\documentclass[12pt]{article}

\usepackage{scicite}

\usepackage{times}

\usepackage{url}
\usepackage{verbatim}
\usepackage{rotating}

\expandafter\def\expandafter\UrlBreaks\expandafter{\UrlBreaks
  \do\a\do\b\do\c\do\d\do\e\do\f\do\g\do\h\do\i\do\j%
  \do\k\do\l\do\m\do\n\do\o\do\p\do\q\do\r\do\s\do\t%
  \do\u\do\v\do\w\do\x\do\y\do\z\do\A\do\B\do\C\do\D%
  \do\E\do\F\do\G\do\H\do\I\do\J\do\K\do\L\do\M\do\N%
  \do\O\do\P\do\Q\do\R\do\S\do\T\do\U\do\V\do\W\do\X%
  \do\Y\do\Z}

\newcommand{\picks}[1]{{\em #1}}

\newcommand{\mt}[1]{\ensuremath{\mathtt{#1}}}

\newenvironment{sciabstract}{%
\begin{quote} \bf}
{\end{quote}}

\newcounter{lastnote}

\title{An Investigation of COVID-19 Spreading Factors with Explainable AI Techniques}

\author
{Xiuyi Fan,$^{1\ast}$ Siyuan Liu,$^{1}$ Jiarong Chen$^{2,3,4},$ Thomas C. Henderson$^{5}$\\
\\
\normalsize{$^{1}$Computer Science Department, Swansea University, United Kingdom}\\
\normalsize{$^{2}$Clinical Experimental Center, Jiangmen Central
  Hospital, China}\\
\normalsize{$^{3}$Department of Oncology, Jiangmen Central Hospital,
  China}\\
\normalsize{$^{4}$Computational Oncology Group, Imperial College London, United Kingdom}\\
\normalsize{$^{5}$School of Computing, University of Utah, USA}\\
\\
\normalsize{$^\ast$E-mail: xiuyi.fan@swansea.ac.uk}
}

\date{}

\begin{document}


\baselineskip24pt


\maketitle

\begin{sciabstract}

BACKGROUND:

\noindent
Since COVID-19 was first identified in December 2019, various
public health interventions have been implemented across the world. As
different measures are implemented at different countries at different
times, we conduct an assessment of the relative effectiveness of the
measures implemented in 18 countries and regions using data
from 22/01/2020 to 02/04/2020.

METHODS:

\noindent
We compute the top one and two measures that are most effective
for the countries and regions studied during the period. Two
Explainable AI techniques, SHAP and ECPI, are used in our study; such
that we construct (machine learning) models for predicting the
instantaneous reproduction number ($R_t$) and use the models as
surrogates to the real world and inputs that the greatest influence to
our models are seen as measures that are most effective.

FINDINGS:

\noindent
Across-the-board, 
city lockdown and contact tracing are the two most effective
measures. For ensuring $R_t<1$, public wearing face masks is also
important. Mass testing alone is not the most effective measure
although when paired with other measures, it can be effective. Warm
temperature helps for reducing the transmission.

INTERPRETATION:

\noindent
After a period of city lockdown, the transmission of COVID-19 has been
slowed. However, as countries are considering lifting city lockdown in
the next a few weeks or months, to prevent resurgent disease, effort
should be put to developing privacy preserving, practical and
effective contact tracing techniques.

FUNDING:

\noindent
The first author is partly supported by the EPSRC CHERISH Digital
Economy Centre.
The 3rd author is supported by the Guangdong
International Young Research Talents Training Programe for
Postdoctoral Researcher.
\end{sciabstract}

\section*{Research in Context}

\paragraph*{Evidence before this study}

Since COVID-19 was first identified in December 2019, various public
health interventions have been implemented across the world. Wuhan has
been locked down since Jan 23, 2020; France closed its schools on
March 16, 2020; South Korea banned international travelers from
Hubei on Feb 02, 2020; Singapore started contact tracing on January
23, 2020, etc. As for mid of April 2020, it seems countries that have
implemented some of these measures are seeing a reduced rate of number
of confirmed cases growth. Yet, the different countermeasures
implemented by different countries at different time still pose the
questions: {\em which countermeasures are effective?} Moreover, the
effectiveness of mass use of face masks has been a controversial
issue, with claims that mass use face mask being ineffective and
essential both exist. Lastly, the impact of temperature and humidity
is also not entirely clear. Thus, we present a comprehensive study
that examines data from 18 countries and regions to understand the
effectiveness of such measures. We have used the {\em Time Series
  Covid 19 Confirmed} data set from
\url{https://www.kaggle.com} for the number of confirmed cases, {\em
  COVID 19 Containment measures} data set also from
\url{https://www.kaggle.com} together with Wikipedia and Google search
with keywords containing country name and control measures such as
``South Korea International Travel Ban'' to find the dates of measure
implementation, and data from
\url{https://rp5.ru/Weather_in_the_world} to find the temperature and
humidity.

\paragraph*{Added value of this study}

As of mid-April 2020, most countries in eastern Europe and Africa are
still in the early stage of the pandemic, with relatively small number
of confirmed cases; understanding the effectiveness of countermeasures
are important for controlling the spread. Countries in western Europe,
Australasia and North America which have implemented strict control
measures for a few weeks are discussing relaxing such measures
including lifting lockdown; understanding the effectiveness of
measures could help governments in these countries adjust their
policies so they do not risk resurgent disease while saving healthcare
and social resources.

\paragraph*{Implications of all the available evidence}

We identify city lockdown and contact tracing as the most effective
countermeasures. Mass testing, mask use by the public, and warm
weather are also useful for controlling transmission. For countries
that are implementing city lockdown, once that is lifted,
contact tracing has been most successfully implemented with technology
based approaches such as tracking mobile phones, developing policies
along with technologies that support such contact tracing while
providing privacy protection should be seriously considered; promoting
mask use and ensuring its supplies should also be considered.

\section*{Introduction}



COVID-19 has spread globally for more than three months since it was
first identified in December 2019. Shortly after its initial
identification, various control measures have been implemented in
different countries for the purposes of containing and delaying the
pandemic; at the moment, some of the less affected countries are
considering implementing these measures. Thus, a major question to be
answered is that: {\em among these control measures, what are the more
  effective ones, at different stages of the disease development?}

Thus, the overarching goal of this work is to identify factors that
are most influential in controlling the spread of the disease. $R_t$,
the average number of secondary cases generated by one primary case
with symptom onset on day $t$, is one of the most important quantities
used to measure the epidemic spread. If $R_t>1$ , then the epidemic is
expanding at time $t$, whereas if $R_t<1$, then it is shrinking at
time $t$. We want to identify factors that are most influential for
controlling $R_t$.

Explainable AI (XAI) is a rising field in AI. In addition to
developing AI systems that making accurate predictions, XAI
systems ``explain'' their predictions \cite{Miller19,Biran17,Doran17b,Elenberg17,Ribeiro18,Wachter2017}.
The development of XAI is motivated by building
trustworthy systems and revealing insights from data. Still in its
infancy, several XAI techniques such as Shapley additive explanations
(SHAP) \cite{Lundberg17} and Explainable Classification with
Probabilistic Inferences (ECPI) \cite{Fan2020}, have been developed in
recent years for identifying {\em decisive features} in prediction
tasks. These techniques are data-driven; they ``explain'' a
prediction by pointing out factors that are ``most significant'' for
the prediction purely based on the data provided.

Much effort has been put into COVID-19 data collection by the global
community. As of early April 2020, time series data containing daily
confirmed cases in more than 100 countries are made publicly available
at online repositories. With help of Internet search engines, one can
identify control measures implemented in different countries and their
respective implementation time. Using such data, with {\em Machine
  Learning (ML)} techniques, we construct classification models that
{\em predict} $R_t$.
We then apply XAI techniques to examine the developed ML models and
identify key factors that are most influential to their
predictions. In this way, the constructed ML models serve as
surrogates to the real world; and identifying effective factors in
controlling $R_t$ becomes explaining the classifications given by the
developed ML models.

\begin{sidewaystable}
\centering
\caption{Implementation dates of control measures at 18 countries and
  regions.\label{table:main}}
\begin{footnotesize}
\begin{tabular}{|c|ccccccc|}
\hline
Countries and & Government      & Mask       & School       & City          & Mass         & International    & Contact \\
Regions       & Advocation (GA) & Use (MU)   & Closure (SC) & Lockdown (CL) & Testing (MT) & Travel Ban (ITB) & Tracing (CT) \\
\hline
Australia   & 13/03/2020        &            &              &               &              & 01/02/2020       & \\
France      & 12/03/2020        &            & 16/03/2020   & 17/03/2020    &              & 16/03/2020       & \\
Germany     & 28/01/2020        &            & 26/02/2020   & 16/03/2020    &              & 28/01/2020       & \\
Italy       & 31/01/2020        &            & 04/03/2020   & 08/03/2020    &              & 31/01/2020       & \\
Japan       & 24/01/2020        & 22/01/2020 & 02/03/2020   &               &              & 01/02/2020       & 25/02/2020 \\
Singapore   & 22/01/2020        & 01/02/2020 &              &               & 24/01/2020   & 29/01/2020       & 23/01/2020 \\
South Korea & 22/01/2020        & 22/01/2020 & 22/01/2020   &               & 31/01/2020   & 02/02/2020       & 22/01/2020 \\
Spain       & 14/03/2020        &            & 12/03/2020   & 14/03/2020    &              & 10/03/2020       & \\
United Kingdom & 01/03/2020     &            & 20/03/2020   & 21/03/2020    &              &                  & \\
\hline
Beijing     & 24/01/2020        & 07/02/2020 & 22/01/2020   & 24/01/2020    & 24/01/2020   & 28/03/2020       & 24/01/2020\\
California  & 04/03/2020        &            & 13/03/2020   & 19/03/2020    &              & 02/02/2020       & \\
Guangdong   & 23/01/2020        & 26/01/2020 & 22/01/2020   & 24/01/2020    & 23/01/2020   & 28/03/2020       & 23/01/2020  \\
Hong Kong   & 04/01/2020        & 08/01/2020 & 22/01/2020   &               & 04/01/2020   & 27/01/2020       & 04/01/2020 \\
Hubei       & 20/01/2020        & 22/01/2020 & 22/01/2020   & 23/01/2020    & 05/02/2020   & 23/01/2020       & 03/02/2020 \\
Macau       & 31/12/2019        & 03/02/2020 & 22/01/2020   &               & 20/02/2020   & 28/01/2020       & \\
New York    & 07/03/2020        &            & 15/03/2020   & 20/03/2020    & 13/03/2020   & 02/02/2020       & \\
Taiwan      & 20/01/2020        & 31/01/2020 & 22/01/2020   &               & 01/02/2020   & 23/01/2020       & 27/01/2020 \\
Washington  & 29/02/2020        &            & 13/03/2020   & 23/03/2020    & 17/03/2020   & 02/02/2020       & \\
\hline
\end{tabular}
\end{footnotesize}
\end{sidewaystable}

\section*{Methods}

\paragraph*{Data Preparation.}

Our analysis is based on the following information:

\begin{itemize}

\item
Implementation dates of control measures as shown in
Table~\ref{table:main}.

\item
The daily reported number of confirmed cases from 22/01/2020 to
02/04/2020 in countries and regions shown in Table~\ref{table:main}.

\item
Temperature and humidity during our study period at these countries
and regions.

\end{itemize}
From daily reported number of confirmed cases, we apply a method of
estimating $R_t$ as reported in Flaxman, et al. \cite{Flaxman20}.
We start by introducing a {\em serial interval distribution}, which
models the time between a person getting infected and he/she
subsequently infecting another people, as a {\em Gamma}
distribution $\mt{g}$ with mean 7 and standard deviation 4.5 (these
two parameters are reported in \cite{wu2020estimating}); we also
assume this serial interval distribution is the same
for all countries and regions studied in this work. The number of new
infections $c_{t}$ on a given day $t$ is given by
the following discrete convolution function:
\begin{equation}
\label{eqn:1}
c_{t} = R_{t} \sum_{\tau=0}^{t-1}c_{\tau} g_{t-\tau},
\end{equation}
where $g_s = \int_{\tau=s-0.5}^{s+0.5} \mt{g}(\tau)d\tau$ for $s =
2,3,\ldots$ and $g_1 = \int_{\tau=0}^{1.5} \mt{g}(\tau)d\tau$,
$c_{\tau}$ is the number of new infections on day $\tau$. Thus,
new infections identified on day $t$ depend on the number of new
infections in days prior to $t$, weighted by the discretized serial
interval distribution, which is the aforementioned Gamma distribution.

From Equation~\ref{eqn:1}, solve for $R_{t}$, we have:
\begin{equation}
\label{eqn:2}
R_t = \frac{c_t}{\sum_{\tau=0}^{t-1}c_\tau g_{t-\tau}}
\end{equation}
Since $c_t$ and $c_\tau$ are available from our data directly,
e.g., $c_t$ is the difference between the confirmed case on day
$t$ and the confirmed case on day $t-1$,
and $g_{t-\tau}$ can be obtained by integrating the Gamma
distribution, we now have a way to compute $R_t$ for all countries on
all days between 22/01/2020 and 02/04/2020.

\begin{figure}
\begin{center}
\includegraphics[width=\textwidth]{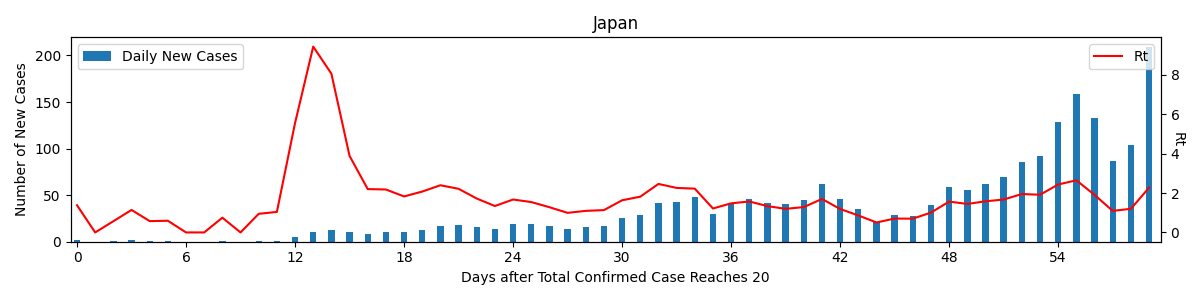}
\includegraphics[width=\textwidth]{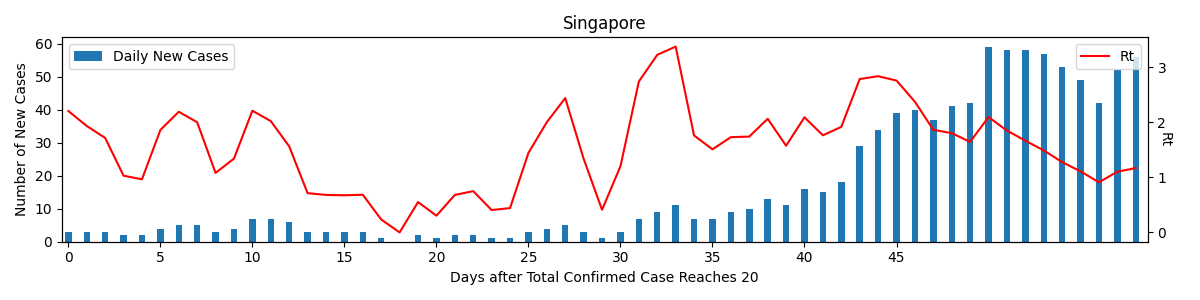}
\includegraphics[width=\textwidth]{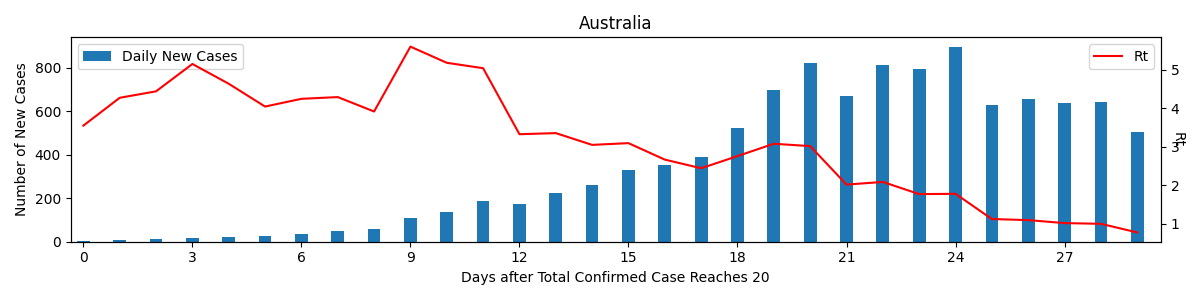}
\includegraphics[width=\textwidth]{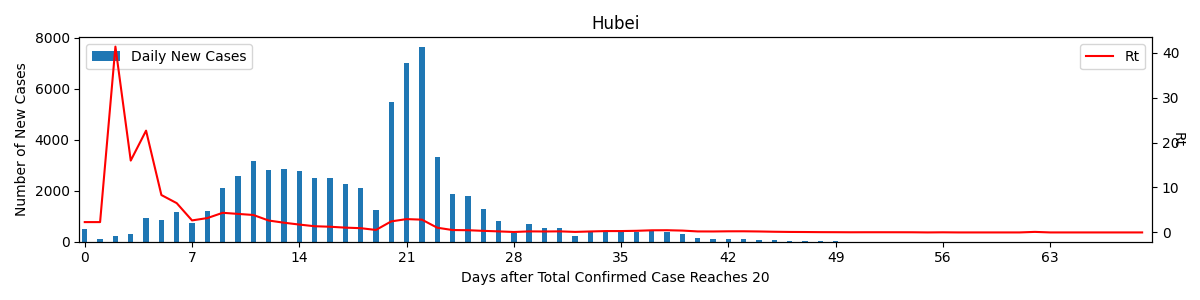}
\end{center}
\caption{Daily new cases and computed $R_t$ for selected countries and
  regions for illustration.}\label{fig:NCRT}
\end{figure}

We then compose a data set in a tabular form where each row describes
information for one country/region on a day, containing the number of
new confirmed cases on that day, days since each of the control
measures that have been implemented, and the temperature as well as
humidity of that day. $R_t$ is added to every row in the data set
and later used as the target for prediction. Since $R_t$ calculated as
such is sensitive to noise at the number of new infection cases on a
given day and the data set we use contains imperfection, e.g., for the
United Kingdom, both 14/03/2020 and 15/03/2020 have 1140 confirmed
cases so there is no increase on 15/03/2020, we thus run a
sliding-window mean filter with radius 1 on the data for noise removal.
Moreover, as $R_t$ calculated in Equation~\ref{eqn:2} assumes a
reasonably large $t$, (otherwise both $c_{\tau}$ and $g_{t-\tau}$
would be too small, resulting an artificially large $R_t$), we drop
entries with confirmed case less than 20. In other words, we only use
data where there are more than 20 accumulated confirmed cases in that
country / region; and as the number of confirmed case is monotonically
increasing, there is no ``skipped'' dates. For instance, our Singapore
cases start on 03/02/2020, Japan cases start on 02/02/2020, and
Germany cases start on 25/02/2020. Figure~\ref{fig:NCRT} illustrates
$R_t$ computed by the presented method for Japan, Singapore, Australia
and Hubei.

A fraction of this data set is shown in Table~\ref{table:illustration}
for illustration. The data in the first row records that $R_t$
is 1.34, with 4 new confirmed cases, 22 days after government
advocation, 12 days after mass use of face mask, school
closure, and city lockdown not implemented,
20 days after mass testing, 15 days after international travel
ban, 21 days after contract tracing, as well as
temperature and humidity 27.825 and 74.37, respectively. The data set
contains 800 entries in total.

\begin{table}[h]
\caption{An illustration of the data set with four data entries
  (Singapore, 12/02/2020, Japan, 26/03/2020, Germany,
  26/03/2020, South Korea, 16/03/2020, and Guangdong, 08/02/2020). NC,
  GA, MU, SC, CL, MT, ITB, CT, T and H are shorthand for New Case,
  Government Advocation, Mask Use, School Closure, City Lockdown,
  International Travel Ban, Contact Tracing, Temperature and Humidity,
  respectively.
\label{table:illustration}}
\centerline{
\begin{tabular}{|c|cccccccccc|}
\hline
$R_t$ & NC  & GA & MU & SC & CL  & MT & ITB & CT & T & H \\
\hline
1.34 & 4    & 22 & 12 & 0  & 0  & 20 & 15 & 21 & 27.86 & 83.86 \\
1.91 & 92   & 63 & 65 & 25 & 0  & 0  & 55 & 31 & 17.375 & 32.75 \\
2.14 & 5962 & 59 & 0  & 30 & 11 & 0  & 12 & 0  & 6.19  & 39.35 \\
0.31 & 78   & 55 & 55 & 55 & 0  & 46 & 44 & 55 & 3.73  & 48.47 \\
0.72 & 53   & 17 & 14 & 18 & 16 & 17 & 0  & 18 & 15.89 & 62.66 \\
\hline
\end{tabular}
}
\end{table}

Since we aim for obtaining easily interpretable qualitative results,
we further discretize our data into category intervals as follows:

\begin{itemize}
\item
NC: [0,10), [10, 100), [100, $\infty$)

\item
GA, MU, SC, CL, MT, ITB, CT: [0,1), [1, 5), [5, 10), [10,15),
      [15, $\infty$)

\item
T: (-$\infty$, 0), [0, 10), [10, 20), [20, $\infty$)

\item
H: [0, 40), [40, 80), [80, $\infty$)
\end{itemize}
\noindent
Thus, the number of new cases is put into 3 categories, represented
with integers 0\ldots2, respectively.
Each of GA, MU, SC, CL,
PTC, MT, ITB and CT is discretized into 5 categories, with each
category represented with an integer 0\ldots4. Similarly,
temperature and humidity are discretized into 4 and 3 categories,
respectively. Table~\ref{table:disc} shows the result of
discretization from data shown in Table~\ref{table:illustration}.

\begin{table}[h]
\caption{Five data entries in Table~\ref{table:illustration} after
  discretization. \label{table:disc}}
\centerline{
\begin{tabular}{|c|cccccccccc|}
\hline
$R_t$ & NC & GA & MU & SC & CL & MT & ITB & CT & T & H \\
\hline
1.34 & 0   & 4  & 3  & 0  & 0  & 4  & 3	  & 4  & 3 & 2 \\
1.91 & 1   & 4  & 4  & 4  & 0  & 0  & 4   & 4  & 2 & 0 \\
2.14 & 2   & 4  & 0  & 4  & 3  & 0  & 3   & 0  & 1 & 0 \\
0.31 & 1   & 4	& 4  & 4  & 0  & 4  & 4   & 4  & 1 & 1 \\
0.72 & 1   & 4  & 3  & 4  & 4  & 4  & 0   & 4  & 2 & 1 \\
\hline
\end{tabular}
}
\end{table}

\paragraph*{XAI Techniques and Problem Formulation.}
Two different XAI techniques for identifying decisive features have
been used in this study: Shapley additive explanations (SHAP), and
Explainable Classification with Probabilistic Inferences (ECPI).
With both SHAP and ECPI, we formulate the factor importance problem as
two binary {\em Explainable Classification} problems: given a data
entry (row) as shown in Table~\ref{table:disc}, classify whether the
$R_t$ (for that row) is greater than some threshold $\theta$; if an
entry is classified as negative ($R_t$ thus less than or equal to
$\theta$), then identify the features that are most influential to the
classification.

SHAP is based on Shapley value, a game theory concept that assigns a
unique distribution of a total surplus generated by the coalition of
all players in a cooperative game. In our context, each factor with
its value, e.g., NC = 0, GA = 4, MU = 3, etc., is viewed as a
``player'' in the game where the outcome is in one of the two classes
($R_t$ being either greater than $\theta$, or not). Shapley value for
each feature-value describes its
``contribution'' to the outcome classification. ECPI is a probabilistic
logic based explainable classification algorithm. ECPI maps a dataset
into a knowledge-base in probabilistic logic and performs
classification with probabilistic logic inference. ECPI computes
explanation by identifying the subset of feature values that gives the
same inference result as the full set (or as close as possible to the
full set when it is not possible to infer the same result). In our
context, roughly speaking, from feature value NC = 0, GA = 4, MU = 3,
etc., one deduces $R_t$ is in some class (either $R_t < \theta$ or
not), then the subset of these feature values that also infer $R_t$ in
the same class is the explanation.

Both SHAP and ECPI identify key features for prediction instances with
SHAP being a model-agnostic method that computes only feature
importance and ECPI an interpretable model that makes the prediction as
well. Technically, two major differences between SHAP and ECPI are that
(1) SHAP considers features individually when evaluating their
``contribution'' whereas ECPI considers all coalition among
features; (2) SHAP estimates the Shapley characteristic function, the
function describing the total expected sum of payoffs the players can
obtain by cooperation, from data whereas ECPI does not perform this
estimation.

Since our goal is to obtain explanations for cases where $R_t<\theta$,
the entire dataset is used for training the ML models. Then, for each
entry in the dataset which has its $R_t < \theta$, we compute the top
$k$ most influential features with both SHAP and ECPI. For instance,
for the entry (Guangdong, 08/02/2020):

\centerline{
\begin{tabular}{|cccccccccc|}
\hline
NC & GA & MU & SC & CL & MT & ITB & CT & T & H \\
\hline
1  & 4  & 3  & 4  & 4  & 4  & 0   & 4  & 2 & 1 \\
\hline
\end{tabular}
}
\noindent
we let $\theta = 1$, then: SHAP finds CT=4 (CT implemented for more
than 15 days) and CL=4 (CL implemented for more than 15 days) being
the most and second most influential factors, respectively; whereas
ECPI finds CL=4 and MU=3 (MU implemented for 10-15 days) being the top
two factors. The identified most influential features from all entries
are aggregated and reported in the next section.

\section*{Results}

\begin{figure}
\begin{center}
\includegraphics[width=\textwidth]{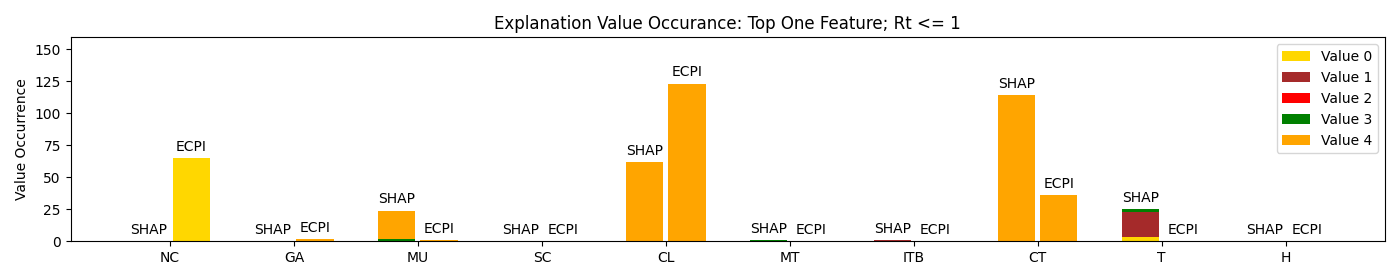}
\includegraphics[width=\textwidth]{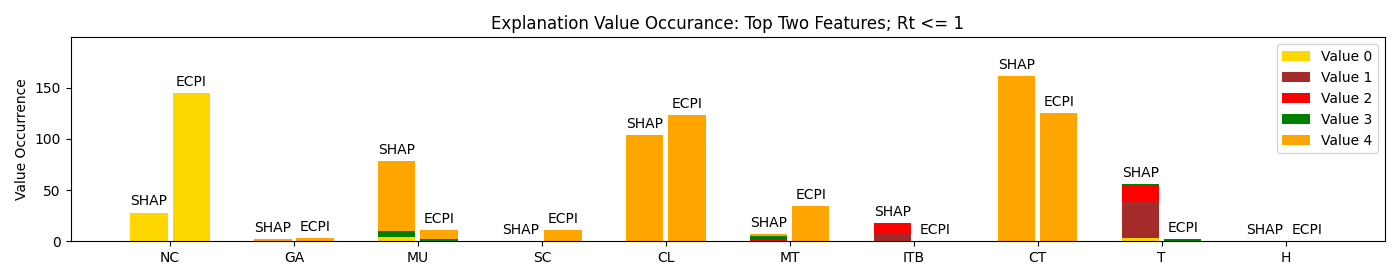}
\includegraphics[width=\textwidth]{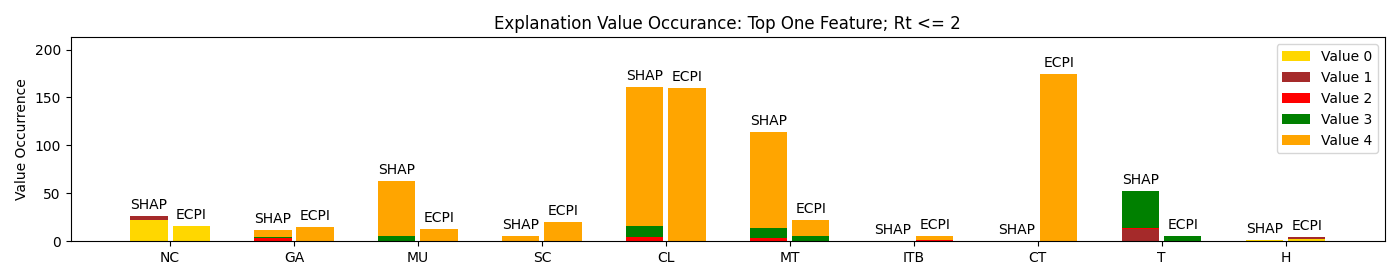}
\includegraphics[width=\textwidth]{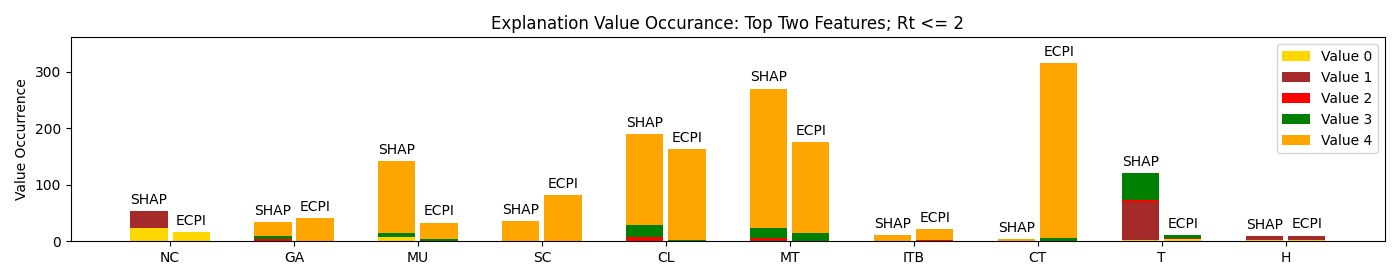}
\end{center}
\caption{Top $k=1,2$ influential factors for $R_t < 1$ and $R_t <
  2$.\label{fig:main}}
\end{figure}

With SHAP and ECPI, we study the two classification cases for
$\theta=1,2$. For each case, we compute the top $k$ ($k = 1,2$)
features that are the most influential. There are 228 and 435 entries
with $R_t < 1$ and $R_t < 2$, respectively. The results are shown in
Figure~\ref{fig:main}.

From Figure~\ref{fig:main}, several qualitative interpretations can be
obtained. We can see that \picks{City Lockdown (CL)}
and \picks{Contact Tracing (CT)} are most influential, for all
parameter combinations ($\theta=1,2$ and $k=1,2$). Both measures are
effective when they take value 4, meaning they are implemented for
more than 15 days. \picks{Mask Use (MU)} and \picks{Mass Testing (MT)}
both show some influence, although not to the level of \picks{CL} and
\picks{CT}. \picks{School Closure (SC)}, \picks{International Travel
  Ban} and \picks{Government Advocation (GA)} rank lower in their
effectiveness. Humidity plays no role completely whereas warm
temperature could be helpful. The bars shown on \picks{New Cases (NC)}
might be interpreted as: when $R_t$ is sufficiently small, it is
likely to stay in that case.

\begin{figure}
\begin{center}
\includegraphics[width=0.49\textwidth]{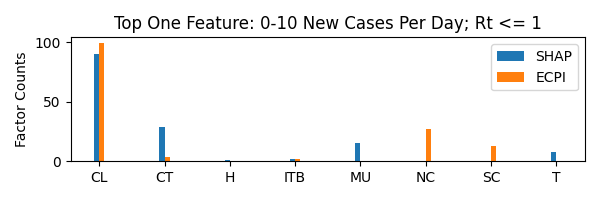}
\includegraphics[width=0.49\textwidth]{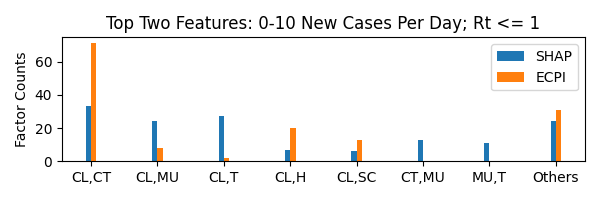}
\includegraphics[width=0.49\textwidth]{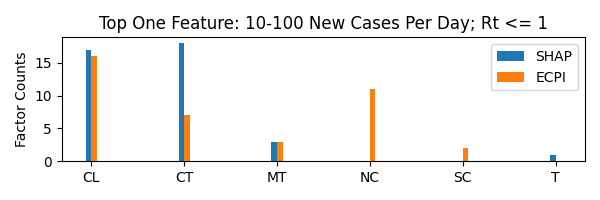}
\includegraphics[width=0.49\textwidth]{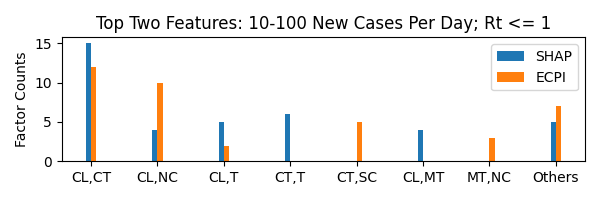}
\includegraphics[width=0.49\textwidth]{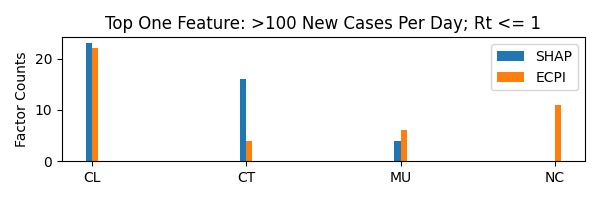}
\includegraphics[width=0.49\textwidth]{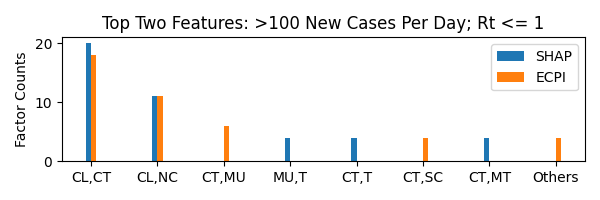}
\end{center}
\caption{Influential factors for daily new cases in different
  ranges.}\label{fig:sub1}
\end{figure}

\begin{figure}
\begin{center}
\includegraphics[width=0.49\textwidth]{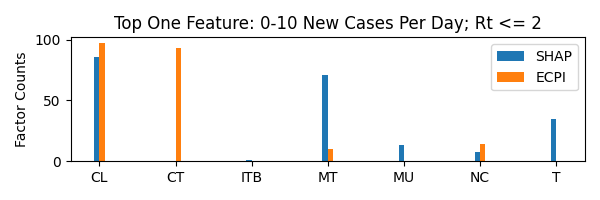}
\includegraphics[width=0.49\textwidth]{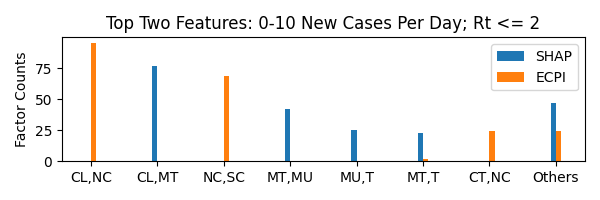}
\includegraphics[width=0.49\textwidth]{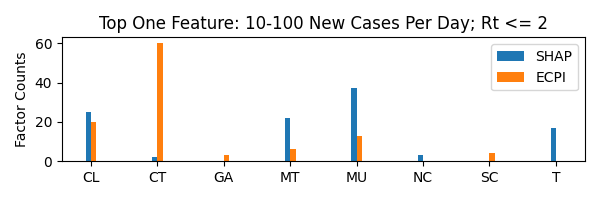}
\includegraphics[width=0.49\textwidth]{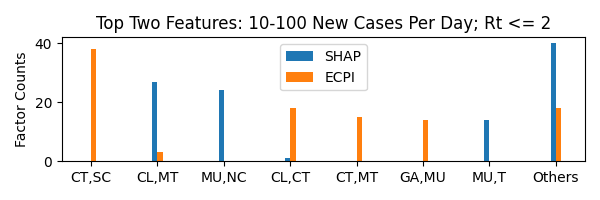}
\includegraphics[width=0.49\textwidth]{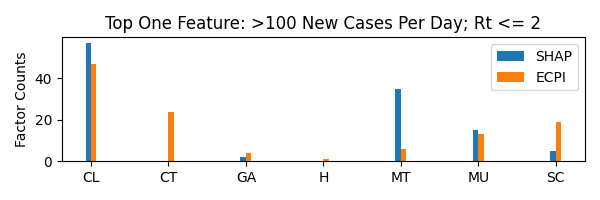}
\includegraphics[width=0.49\textwidth]{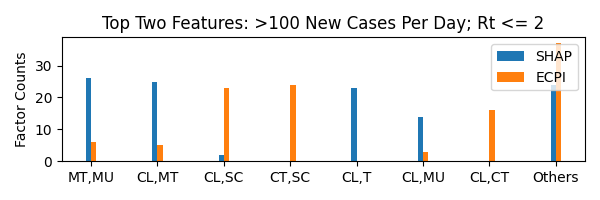}
\end{center}
\caption{Influential factors for daily new cases in different
  ranges ($R_t < 2$).}\label{fig:sub2}
\end{figure}

Figure~\ref{fig:sub1} and \ref{fig:sub2} show influential factors for
daily new cases in three ranges: 0-10, 10-100, and $>100$ for $R_t <
1$ and $R_t < 2$, respectively. For $R_t<1$, we see that \picks{CL} is
the most effective single measure and \picks{CL,CT} the most effective
pair, regardless the number of new cases, agreed by both SHAP and
ECPI. For $R_t < 2$, the top effective measures are \picks{CL},
\picks{CT}, \picks{MT} and \picks{MU}. The top pairs are harder to
select, \picks{CL,NC}, \picks{CL,MT}, \picks{CT,SC} and \picks{MT,MU}
are more effective than others.

\section*{Discussion}

Since COVID-19 was identified in December 2019, there are a few works
which have been dedicated to understanding the effect of non-pharmaceutical
countermeasures. Leung et al. \cite{leung2020first}
estimated $R_{t}$ in four Chinese cities and ten province between
mid January to mid February. They found that though aggressive
non-pharmaceutical interventions such as city lockdown have made the
first wave of COVID-19 outside of Hubei abated, control measures should
be relaxed gradually. Our results, asserting city lockdown being the
most effective measure confirms their findings.

In~\cite{ng2020evaluation}, the authors evaluated the effectiveness of
surveillance and containment measures for the first 100 patients with
COVID-19 in Singapore. The surveillance strategy in
Singapore includes applying the case definition at medical consultations,
tracing contacts of patients, enhancing surveillance among different
patient groups and allowing clinician discretion. It was found that
rapid identification and isolation of cases, quarantine of close
contacts, and active monitoring of other contacts have been effective
in suppressing expansion of the outbreak. Our results find that
contact tracing being the overall second effective measure (after city
lockdown) confirms their results.

In \cite{kraemer2020effect}, the authors studied the real-time
mobility data from Wuhan and detailed case data including travel
history to investigate the role of travel restrictions in limiting the
spread of COVID-19. It was found that travel restrictions are
particularly useful in the early stage of an outbreak, while the
effect will drop when the outbreak is more widespread. A combination
of interventions may be necessary though the individual role of each
intervention is yet determined. Our results show that international
travel ban, although exactly same as their work on inter-city travel,
but also on the effectiveness of restricting travel, is not nearly as
effective as other measures such as city lockdown and contact
tracing.

In~\cite{prem2020effect}, the authors built an age-structured
susceptible-exposed-infected-removed (SEIR) model to estimate the
effect of physical distancing measures, such as extended school
closures and workplace distancing, on the progression of the COVID-19
epidemic. It was found that sustaining these measures are effective in
reducing the size of epidemic. Our results, especially the ones from
ECPI, show that school closure has a positive effect in lowering
$R_t$. 

As a data-driven modeling approach, our work is limited by a number
of factors. Firstly, all results are based on data collected from the
selected 18 countries and regions during the period of 22/01/2020 to
03/04/2020, although results might be generalizable, but are about
these regions during the said period. Thus, when applying these
results to other regions and other time, they should be viewed
indicative. Secondly, the data used is inherently ambiguous, e.g.,
``contact tracing'' and ``mass testing'' have been implemented at
different countries, but it is unlikely the same standard has been
applied. Thus, although our methods are quantitative,
due to the qualitative nature of the data, one should read our results
qualitatively. Thirdly, we rely on the calculated $R_t$ to label our
data, which is then used to construct our models. The calculation used
is reported in \cite{Flaxman20} with parameters found in
\cite{wu2020estimating}. We are aware that $R_t$ is an estimate that
can be approximated with more than one method, some authors such as
\cite{leung2020first} give a much smaller estimate of $R_t$ for
Beijing in January (they estimate $R_t$ being close to 0.5 whereas our
calculation shows it is greater than 2; although ours drops to below
0.5 after February 10, same as theirs), different results might be
obtained if $R_t$ is estimated differently.

In conclusion, we applied two machine learning and explainable AI
methods in studying the influence of factors affecting the
spread of COVID-19. We find city lockdown and contact tracing being
the two most effective control measures, surpassing mass testing,
school closure, international travel ban and mask use. As countries
are considering lifting city lockdown, to prevent resurgent disease,
effort should be put to developing privacy preserving, practical and
effective contact tracing techniques.

\bibliographystyle{acm}
\bibliography{covid19sy}

\begin{thebibliography}{10}

\bibitem{Biran17}
{\sc Biran, O., and Cotton, C.~V.}
\newblock Explanation and justification in machine learning : A survey.
\newblock In {\em Proc. of IJCAI-17 Workshop on Explainable AI\/} (2017).

\bibitem{Doran17b}
{\sc Doran, D., Schulz, S., and Besold, T.~R.}
\newblock What does explainable {AI} really mean? {A} new conceptualization of
  perspectives.
\newblock In {\em Proc. of AI*IA\/} (2017).

\bibitem{Elenberg17}
{\sc Elenberg, E., Dimakis, A., Feldman, M., and Karbasi, A.}
\newblock Streaming weak submodularity: Interpreting neural networks on the
  fly.
\newblock In {\em Proc. of NIPS\/} (2017).

\bibitem{Fan2020}
{\sc Fan, X., Liu, S., and Henderson, T.~C.}
\newblock Explainable ai for classification using probabilistic logic
  inference.
\newblock {\em arXiv\/} (2020).

\bibitem{Flaxman20}
{\sc Flaxman, S., Mishra, S., Gandy, A., Unwin, H., Coupland, H., Mellan, T.,
  Zhu, H., Berah, T., Eaton, J., Perez~Guzman, P., et~al.}
\newblock Report 13: Estimating the number of infections and the impact of
  non-pharmaceutical interventions on covid-19 in 11 european countries.
\newblock Tech. rep., Imperial College London, 2020.

\bibitem{kraemer2020effect}
{\sc Kraemer, M.~U., Yang, C.-H., Gutierrez, B., Wu, C.-H., Klein, B., Pigott,
  D.~M., du~Plessis, L., Faria, N.~R., Li, R., Hanage, W.~P., et~al.}
\newblock The effect of human mobility and control measures on the covid-19
  epidemic in china.
\newblock {\em Science\/} (2020).

\bibitem{leung2020first}
{\sc Leung, K., Wu, J.~T., Liu, D., and Leung, G.~M.}
\newblock First-wave covid-19 transmissibility and severity in china outside
  hubei after control measures, and second-wave scenario planning: a modelling
  impact assessment.
\newblock {\em The Lancet\/} (2020).

\bibitem{Lundberg17}
{\sc Lundberg, S.~M., and Lee, S.}
\newblock A unified approach to interpreting model predictions.
\newblock In {\em Proc. of NIPS\/} (2017).

\bibitem{Miller19}
{\sc Miller, T.}
\newblock Explanation in artificial intelligence: Insights from the social
  sciences.
\newblock {\em AI Journal\/} (2019).

\bibitem{ng2020evaluation}
{\sc Ng, Y., Li, Z., Chua, Y.~X., Chaw, W.~L., Zhao, Z., Er, B., Pung, R.,
  Chiew, C.~J., Lye, D.~C., Heng, D., et~al.}
\newblock Evaluation of the effectiveness of surveillance and containment
  measures for the first 100 patients with covid-19 in singapore--january
  2--february 29, 2020.

\bibitem{prem2020effect}
{\sc Prem, K., Liu, Y., Russell, T.~W., Kucharski, A.~J., Eggo, R.~M., Davies,
  N., Flasche, S., Clifford, S., Pearson, C.~A., Munday, J.~D., et~al.}
\newblock The effect of control strategies to reduce social mixing on outcomes
  of the covid-19 epidemic in wuhan, china: a modelling study.
\newblock {\em The Lancet Public Health\/} (2020).

\bibitem{Ribeiro18}
{\sc Ribeiro, M.~T., Singh, S., and Guestrin, C.}
\newblock Anchors: High-precision model-agnostic explanations.
\newblock In {\em Proc of AAAI-18\/} (2018).

\bibitem{Wachter2017}
{\sc Wachter, S., Mittelstadt, B., and Floridi, L.}
\newblock Transparent, explainable, and accountable ai for robotics.
\newblock {\em Science Robotics\/} (2017).

\bibitem{wu2020estimating}
{\sc Wu, J.~T., Leung, K., Bushman, M., Kishore, N., Niehus, R., de~Salazar,
  P.~M., Cowling, B.~J., Lipsitch, M., and Leung, G.~M.}
\newblock Estimating clinical severity of covid-19 from the transmission
  dynamics in wuhan, china.
\newblock {\em Nature Medicine\/} (2020), 1--5.

\end{thebibliography}

\section*{Appendix}

\paragraph*{Classification Performance}

Since the reliability of our explanation results depends on the
quality of our ML models, which depend on the quality of data and the
discretization process we used, we first present some classification
results. The two datasets are randomly divided into 90\% / 10\% split
with 90\% used for training and 10\%
for testing. We present results using a random forest classifier with
100 trees, neural network with two hidden layers with 12 and 10 nodes
each, and ECPI. We measure performance with precision and recall,
which are defined as
\[
precision = \frac{true \ positive}{true \ positive + false \ positive},
\]
and
\[
recall = \frac{true \ positive}{true \ positive + false \ negative}.
\]
The precision and recalls are shown in
Table~\ref{table:precision_recall}. These results show that
effective models can be built with the dataset and our ECPI model
gives indicative predictions.

\begin{table}[h]
\caption{Classification performance with Random Forest (RF), Neural Network (NN) and
  ECPI.\label{table:precision_recall}}
\centerline{
\begin{tabular}{|l|ccc|ccc|}
\hline
          & \multicolumn{3}{c|}{$R_t < 1$} & \multicolumn{3}{c|}{$R_t < 2$} \\
\cline{2-7}
          & RF   & NN    & ECPI & RF    & NN   & ECPI \\
\hline
Precision & 0.79 & 0.81  & 0.88 & 0.66  & 0.91 & 0.91 \\
Recall    & 0.79 & 0.77  & 0.94 & 0.77  & 0.69 & 0.79 \\
\hline
\end{tabular}
}
\end{table}

\paragraph*{Data Source}

\begin{itemize}
\item
Australia

\begin{itemize}
\item
International Travel Ban (ITB): 01/02/2020
\begin{itemize}
\item
On February 1, Australia banned the entry of foreign nationals
from mainland China, and ordered its own returning citizens from China
to self-quarantine for 14 days.

\item
\url{https://en.wikipedia.org/wiki/2020_coronavirus_pandemic_in_Australia}
\item
Pannett Rachel, ``Australia Restricts Travelers From Mainland China as Virus
Impact Spreads", Wall Street Journal. Archived from
the original on  February 25, 2020. Retrieved on March 17, 2020.
\end{itemize}
\end{itemize}

\begin{itemize}
\item
Government Advocation (GA): 13/03/2020
\begin{itemize}
\item
On March 13, the National Cabinet, a form of national crisis cabinet
akin to a war cabinet, was created following a meeting of the Council
of Australian Governments (COAG).

\item
\url{https://en.wikipedia.org/wiki/2020_coronavirus_pandemic_in_Australia\#cite_note-81}
\end{itemize}
\end{itemize}

\item
France

\begin{itemize}
\item
Government Advocation (GA): 12/03/2020

\begin{itemize}
\item
On March 12, French President Emmanuel Macron announced on public
television that all schools and all universities would close from
Monday (March 16) until further notice.

\item
\url{https://en.wikipedia.org/wiki/2020_coronavirus_pandemic_in_France}
\end{itemize}
\end{itemize}

\begin{itemize}
\item
School Closure (SC): 16/03/2020

\begin{itemize}
\item
On March 12, French President Emmanuel Macron announced on public
television that all schools and all universities would close from
Monday (March 16) until further notice.

\item
\url{https://en.wikipedia.org/wiki/2020_coronavirus_pandemic_in_France}
\end{itemize}
\end{itemize}

\begin{itemize}
\item
International Travel Ban (ITB): 16/03/2020

\begin{itemize}
\item
On March 15, France has announced it will be tightening controls on its border
with Germany as of March 16.

\item
\url{https://www.theguardian.com/world/2020/mar/15/coronavirus-causes-french-voters-to-stay-away-from-municipal-elections}
\end{itemize}
\end{itemize}

\begin{itemize}
\item
City Lockdown (CL): 17/03/2020

\begin{itemize}
\item
On March 14, French Prime Minister \'Edouard Philippe ordered the
closure of all non-essential public places, including restaurants,
caf\'es, cinemas, and nightclubs, effective at midnight. On March 16,
President Emmanuel Macron announced mandatory home confinement for 15
days starting at noon on 17 March.

\item
\url{https://en.wikipedia.org/wiki/2020_coronavirus_pandemic_in_France}
\end{itemize}
\end{itemize}

\item
Germany

\begin{itemize}
\item
Government Advocation (GA): 28/01/2020

\begin{itemize}
\item
On January 28, hotlines were established to calm down worried callers. Experts advise unsettled citizens to follow the usual rules that are
also appropriate to protect against a flu infection: regular hand
washing, coughing or sneezing in a handkerchief or keeping your arm
and distance from other sick people.

\item
\url{https://en.wikipedia.org/wiki/2020_coronavirus_pandemic_in_Germany}

\url{https://www.welt.de/wirtschaft/article205424021/Coronavirus-Behoerden-bereiten-sich-auf-hunderte-Infizierte-vor.html}
\end{itemize}
\end{itemize}

\begin{itemize}
\item
School Closure (SC): 26/02/2020

\begin{itemize}
\item
On February 26, following the confirmation of multiple COVID-19 cases
in North Rhine-Westphalia, Heinsberg initiated closure of schools,
swimming pools, libraries and the town hall until March 2.  The closures were
still effective until the day of the data was collected (April 3).

\item
\url{https://en.wikipedia.org/wiki/2020_coronavirus_pandemic_in_Germany}
\end{itemize}
\end{itemize}

\begin{itemize}
\item
International Travel Ban (ITB): 16/03/2020

\begin{itemize}
\item
On March 15, German Interior Minister Horst Seehofer announced to shut down the borders with France, Switzerland, Austria, Denmark and Luxembourg.The measure would begin on Monday (March 16) and the transportation of goods and commuters would be exempt.

\item
\url{https://en.wikipedia.org/wiki/2020_coronavirus_pandemic_in_Germany}
\end{itemize}
\end{itemize}

\begin{itemize}
\item
City Lockdown (CL): 16/03/2020

\begin{itemize}
\item
On March 16, the Germany federal government and the heads of government of the federal states have reached an agreement in view of the corona epidemic in Germany. They decided on guidelines for a uniform approach to further restrict social contacts in the public sector. According to the agreement of the federal and state governments, bars, clubs, discotheques, pubs and similar facilities are to be closed to the public. In addition, theaters, operas, concert halls, museums, trade fairs, exhibitions, cinemas, leisure and animal parks as well as providers of leisure activities (indoors and outdoors) are to shut down. Sports facilities, gyms, swimming pools and fun pools, playgrounds and other retail outlets are also affected.

\item
\url{https://www.bundesregierung.de/breg-de/themen/coronavirus/leitlinien-bund-laender-1731000}
\end{itemize}
\end{itemize}

\item
Italy

\begin{itemize}
\item
Government Advocation (GA): 31/01/2020

\begin{itemize}
\item
On January 31, Italy declared a six-month state of emergency today after two Chinese
tourists tested positive for the coronavirus.

\item
\url{https://www.thetimes.co.uk/article/italy-declares-state-of-emergency-after-two-coronavirus-cases-are-confirmed-r8xlpfqmw}
\end{itemize}
\end{itemize}

\begin{itemize}
\item
International Travel Ban (ITB): 31/01/2020

\begin{itemize}
\item
On January 31, the Italian government suspended all flights to and
from China.

\item
\url{https://en.wikipedia.org/wiki/2020_coronavirus_pandemic_in_Italy}
\end{itemize}
\end{itemize}

\begin{itemize}
\item
School Closure (SC): 04/03/2020

\begin{itemize}
\item
On March 4, the Italian government imposed the shutdown of all schools
and universities nationwide for two weeks as the country reached 100
deaths from the outbreak. The closures were
still effective on the day of the data was collected (April 3).

\item
\url{https://en.wikipedia.org/wiki/2020_coronavirus_pandemic_in_Italy}
\end{itemize}
\end{itemize}

\begin{itemize}
\item
City Lockdown (CL): 08/03/2020

\begin{itemize}
\item
In the night between March 7 and 8, the Italian government approved a decree
to lock down Lombardy and 14 other provinces in Veneto,
Emilia-Romagna, Piedmont and Marche, involving more than 16 million
people. The closures were
still effective on the day of the data was collected (April 3).

\item
\url{https://en.wikipedia.org/wiki/2020_coronavirus_pandemic_in_Italy}
\end{itemize}
\end{itemize}

\item
Japan

\begin{itemize}
\item
Mask Use (MU): 22/01/2020

\begin{itemize}
\item
It is indicted that the use of face masks is ubiquitous in China and other Asian
countries such as South Korea and Japan.

\item
Feng Shuo, Chen Shen, Nan Xia, Wei Song, Mengzhen Fan, and Benjamin J. Cowling. ``Rational use of face masks in the COVID-19 pandemic." \emph{The Lancet Respiratory Medicine} (2020).
\end{itemize}
\end{itemize}

\begin{itemize}
\item
Government Advocation (GA): 24/01/2020

\begin{itemize}
\item
On January 24, Japan Prime Minister Abe convened the "Ministerial Meeting on
Countermeasures Related to the Novel Coronavirus" at the Prime
Minister's Office with members of his Cabinet in response to a
statement released by the World Health Organization (WHO) that morning
which confirmed human-to-human transmission of the
coronavirus.

\item
\url{https://en.wikipedia.org/wiki/2020_coronavirus_pandemic_in_Japan.}
\end{itemize}
\end{itemize}

\begin{itemize}
\item
International Travel Ban (ITB): 01/02/2020

\begin{itemize}
\item
On February 1, Japan Prime Minister Abe announced during the Fourth Meeting of the Novel
Coronavirus Response Headquarters that he would enact restrictions to
deny entry of foreign citizens who had a history of visiting Hubei
province within 14 days and those who possess a Chinese passport
issued by Hubei province.

\item
\url{https://en.wikipedia.org/wiki/2020_coronavirus_pandemic_in_Japan.}
\end{itemize}
\end{itemize}

\begin{itemize}
\item
Contact Tracing (CT): 25/02/2020

\begin{itemize}
\item
On February 25, the Ministry of Health, Labour and Welfare established
a "Cluster Response Section" in accordance to the Basic Policies for
Novel Coronavirus Disease Control. The purpose of the new section is
to quickly identify and contain small-scale clusters of COVID-19
infections before they turn into large-scale ones. It is led by
university professors Oshitani Hitoshi and Nishiura Hiroshi and
consists of a contact trace team and a surveillance team from the
National Institute of Infectious Diseases (NIID), a data analysis team
from Hokkaido University, a risk management team from Tohoku
University, and an administration team.

\item
\url{https://en.wikipedia.org/wiki/2020_coronavirus_pandemic_in_Japan.}
\end{itemize}
\end{itemize}

\begin{itemize}
\item
School Closure (SC): 02/03/2020

\begin{itemize}
\item
On February 27, Japan Prime Minister Abe requested closing all elementary, junior high,
and high schools to curb the spread of the infections from March 2 to
the end of spring vacations, which usually conclude in early
April.

\item
\url{https://en.wikipedia.org/wiki/2020_coronavirus_pandemic_in_Japan}
\end{itemize}
\end{itemize}

\item
Singapore

\begin{itemize}
\item
Government Advocation (GA): 22/01/2020

\begin{itemize}
\item
To combat COVID-19, a multi-ministerial committee was formed on
January 22 with Minister for National Development Lawrence Wong and
Minister for Health Gan Kim Yong as the co-chairs and Prime Minister
Lee Hsien Loong and Deputy Prime Minister and Minister for Finance
Heng Swee Keat as advisors.

\item
\url{https://en.wikipedia.org/wiki/2020_coronavirus_pandemic_in_Singapore}
\end{itemize}
\end{itemize}

\begin{itemize}
\item
Contact Tracing (CT): 23/01/2020

\begin{itemize}
\item
On January 23, The first case in Singapore was confirmed, involving a
66-year-old Chinese national from Wuhan who flew from Guangzhou via
China Southern Airlines flight CZ351 with nine companions. He stayed
at Shangri-La's Rasa Sentosa Resort and Spa. Contact tracing
subsequently commenced.

\item
\url{https://en.wikipedia.org/wiki/2020_coronavirus_pandemic_in_Singapore}
\end{itemize}
\end{itemize}

\begin{itemize}
\item
Mass Testing (MT): 24/01/2020

\begin{itemize}
\item
Temperature screening for travellers arriving at Singapore Woodlands
and Tuas checkpoints began at noon on Friday (January 24).

\item
\url{https://www.channelnewsasia.com/news/singapore/wuhan-virus-woodlands-tuas-checkpoint-temperature-screening-12319724}
\end{itemize}
\end{itemize}

\begin{itemize}
\item
International Travel Ban (ITB): 29/01/2020

\begin{itemize}
\item
Travellers from Hubei were denied entry from noon of
January 29.

\item
\url{https://en.wikipedia.org/wiki/2020_coronavirus_pandemic_in_Singapore}
\end{itemize}
\end{itemize}

\begin{itemize}
\item
Mask Use (MU): 01/02/2020

\begin{itemize}
\item
On February 1, the government distributed four surgical masks to each
household.

\item
\url{https://en.wikipedia.org/wiki/2020_coronavirus_pandemic_in_Singapore}
\end{itemize}
\end{itemize}

\item
South Korea

\begin{itemize}
\item
Government Advocation (GA): 22/01/2020

\begin{itemize}
\item
President Moon Jae-in ordered to prevent the spread of the 2019-nCoV
during Lunar New Year at an council briefing as of January 22.

\item
\url{https://www.cdc.go.kr/board/board.es?mid=a30402000000\&bid=0030\&act=view\&list_no=365844\&tag=\&nPage=14}
\end{itemize}
\end{itemize}

\begin{itemize}
\item
Mask Use (MU): 22/01/2020

\begin{itemize}
\item
It is indicted that the use of face masks is ubiquitous in China and other Asian
countries such as South Korea and Japan.

\item
Feng Shuo, Chen Shen, Nan Xia, Wei Song, Mengzhen Fan, and Benjamin J. Cowling. ``Rational use of face masks in the COVID-19 pandemic." \emph{The Lancet Respiratory Medicine} (2020).
\end{itemize}
\end{itemize}

\begin{itemize}
\item
Contact Tracing (CT): 20/01/2020

\begin{itemize}
\item
The COVID-19 outbreak in China occurred on December 8, 2019, and the
first case in South Korea was reported on January 20, 2020. In the current epidemiological investigation contact investigation
techniques that were used on a limited basis for the Middle East
Respiratory Syndrome (MERS) outbreak in 2015, are being used in all
confirmed cases of COVID-19.

\item
COVID-19 National Emergency Response Center, Epidemiology \& Case Management Team, Korea Centers for Disease Control \& Prevention. ``Contact Transmission of COVID-19 in South Korea: Novel Investigation Techniques for Tracing Contacts." Osong public health and research perspectives vol. 11,1 (2020): 60-63.
\end{itemize}
\end{itemize}

\begin{itemize}
\item
Mass Testing (MT): 31/01/2020

\begin{itemize}
\item
Extended regional and local triage check-up centre for Covid19.

\item
\url{https://www.kaggle.com/paultimothymooney/covid19-containment-and-mitigation-measures/version/19}
\end{itemize}
\end{itemize}

\begin{itemize}
\item
International Travel Ban (ITB): 02/02/2020

\begin{itemize}
\item
On February 2, the South Korea government said it will ban the entry into Korea of any
foreigners who have visited Hubei Province in China within the past
two weeks.

\item
\url{https://www.koreatimes.co.kr/www/nation/2020/02/119_282795.html}
\end{itemize}
\end{itemize}

\begin{itemize}
\item
School Closure (SC): 07/02/2020

\begin{itemize}
\item
According to Korea's JoongAng Ilbo, there are 450 kindergartens, 77 elementary schools, 29 junior high schools, 33 high schools and 3 special schools shut down till February 7, with a total of 592.  on February 23, all kindergartens, elementary schools, middle schools, and high schools were announced to delay the semester start from 2 to 9 March.

\item
\url{https://en.wikipedia.org/wiki/2020_coronavirus_pandemic_in_South_Korea}
\end{itemize}
\end{itemize}

\item
Spain

\begin{itemize}
\item
International Travel Ban (ITB): 10/03/2020

\begin{itemize}
\item
On March 10, the Government of Spain decreed the immediate
cancellation of all direct flights from Italy to Spain until 25
March. The cancellation was still effective on the day of the data was collected (April 3).

\item
\url{https://en.wikipedia.org/wiki/2020_coronavirus_pandemic_in_Spain}
\end{itemize}
\end{itemize}

\begin{itemize}
\item
School Closure (SC): 12/03/2020

\begin{itemize}
\item
On March 12, most of the autonomous communities in Spain shut down their school
systems, initially for two weeks.

\item
\url{https://en.wikipedia.org/wiki/2020_coronavirus_pandemic_in_Spain}
\end{itemize}
\end{itemize}

\begin{itemize}
\item
Government Advocation (GA): 14/03/2020

\begin{itemize}
\item
On March 13, Prime Minister of Spain Pedro S\'anchez announced a
declaration of a nationwide State of Alarm for 15 days, to become
effective the following day after the approval of the Council of
Ministers.

\item
\url{https://en.wikipedia.org/wiki/2020_coronavirus_pandemic_in_Spain}
\end{itemize}
\end{itemize}

\begin{itemize}
\item
City Lockdown (CL): 14/03/2020

\begin{itemize}
\item
On March 13, the Government of the Community of Madrid decreed the shutting down of bars, restaurants and "non-alimentary" shops (only allowing the opening of supermarkets and chemist's shops). On March 14, Asturias, Catalonia, Cantabria, Galicia, Madrid, Murcia and the Basque Country closed all shops except those selling food and basic necessities. The Mayor of Madrid closed parks and public gardens.

\item
\url{https://en.wikipedia.org/wiki/2020_coronavirus_pandemic_in_Spain}
\end{itemize}
\end{itemize}

\item
United Kingdom

\begin{itemize}
\item
Government Advocation (GA): 01/03/2020

\begin{itemize}
\item
By March 1, cases had been
detected in England, Wales, Northern Ireland and
Scotland. Subsequently, Prime Minister Boris Johnson unveiled the
Coronavirus Action Plan, and the government declared the outbreak as
``level 4 incident".

\item
\url{https://en.wikipedia.org/wiki/2020_coronavirus_pandemic_in_the_United_Kingdom}.
\end{itemize}
\end{itemize}

\begin{itemize}
\item
School Closure (SC): 20/03/2020

\begin{itemize}
\item
All schools, nurseries and colleges are to join those in the rest of
the UK in closing on Friday (March 20) ``until further notice", with the only
exception being made for children of key workers and for vulnerable
children.

\item
\url{https://www.theguardian.com/world/2020/mar/18/coronavirus-school-colleges-nurseries-england-close-uk-friday}
\end{itemize}
\end{itemize}

\begin{itemize}
\item
City Lockdown (CL): 20/03/2020

\begin{itemize}
\item
On 20 March, Prime Minister Boris Johnson requested the closure of pubs, restaurants, gyms, entertainment venues, museums and galleries that evening.

\item
\url{https://en.wikipedia.org/wiki/2020_coronavirus_pandemic_in_the_United_Kingdom}
\end{itemize}
\end{itemize}

\item
Beijing

\begin{itemize}
\item
School Closure (SC): 22/01/2020

\begin{itemize}
\item
School Winter Vocation in Beijing was scheduled as from 18/01/2020 to 16/02/2020. The semester start was delayed until the date of the data was collected (April 3).

\item
\url{https://jw.beijing.gov.cn/tzgg/202004/t20200404_1789233.html}
\end{itemize}
\end{itemize}

\begin{itemize}
\item
Government Advocation (GA): 24/01/2020

\begin{itemize}
\item
On January 24, Beijing launched Level 1 Response Mechanism.

\item
\url{https://www.thepaper.cn/newsDetail_forward_5622869}
\end{itemize}
\end{itemize}

\begin{itemize}
\item
City Lockdown (CL): 24/01/2020

\begin{itemize}
\item
In China, Chinese New Year Holiday was scheduled as from 24/01/2020 to 02/02/2020.
On January 31. Beijing government announced that employees for essential business
will resume working on
09/02/2020. Others will remain working from home until April 10.

\item
\url{http://www.gov.cn/zhengce/content/2020-01/27/content_5472352.htm}

\url{http://www.beijing.gov.cn/zhengce/zhengcefagui/202001/t20200131_1622070.html}

\url{http://www.beijing.gov.cn/zhengce/zhengcefagui/202004/t20200410_1799118.html}

\url{http://www.beijing.gov.cn/zhengce/zhengcefagui/202001/t20200129_1621500.html}
\end{itemize}
\end{itemize}

\begin{itemize}
\item
Contact Tracing (CT): 24/01/2020

\begin{itemize}
\item
On January 24, Beijing City Government announced that disease prevention and
control institutions should strengthen the epidemiological
investigation, investigate the source of infection in detail,
determine the spread of the epidemic, assess the impact and possible
development trend of the epidemic, and trace the close
contacts.

\item
\url{http://www.beijing.gov.cn/zhengce/zhengcefagui/202001/t20200128_1621467.html}
\end{itemize}
\end{itemize}

\begin{itemize}
\item
Mass Testing (MT): 24/01/2020

\begin{itemize}
\item
On January 24, Beijing City Government announced that medical institutions
should strengthen the pre-examination and triage work, and guided
patients to special fever outpatient clinics based on their symptoms
and signs and epidemiological history, and refer the confirmed cases
of pneumonia infected by the new coronavirus to designated hospitals
for treatment, and strengthen the prevention and control of nosocomial
infections.

\item
\url{http://www.beijing.gov.cn/zhengce/zhengcefagui/202001/t20200128_1621467.html}.
\end{itemize}
\end{itemize}

\begin{itemize}
\item
Mask Use (MU): 07/02/2020

\begin{itemize}
\item
On February 7, Beijing Municipal People's Congress Standing Committee agreed on the decision to require people to wear masks in public places.

\item
\url{http://bjrb.bjd.com.cn/html/2020-02/08/content_12445565.htm}
\end{itemize}
\end{itemize}

\begin{itemize}
\item
International Travel Ban (ITB): 28/03/2020

\begin{itemize}
\item
On March 26, Chinese government decided to temporarily suspend entry of
foreigners with the current valid visa and residence permit for entry
into China from 0:00 on March 28, 2020.

\item
\url{https://www.fmprc.gov.cn/web/wjbxw_673019/t1761858.shtml}
\end{itemize}
\end{itemize}

\item
California

\begin{itemize}
\item
International Travel Ban (ITB): 02/02/2020

\begin{itemize}
\item
On January 31, the United States imposed an entry ban on all foreign
nationals who were in the People's Republic of China, excluding
Taiwan, Hong Kong, and Macau, in the past fourteen days, effective
5:00 p.m. eastern standard time on February 2, 2020.

\item
\url{https://www.whitehouse.gov/presidential-actions/proclamation-suspension-entry-immigrants-nonimmigrants-persons-pose-risk-transmitting-2019-novel-coronavirus/}
\end{itemize}
\end{itemize}

\begin{itemize}
\item
Government Advocation (GA): 04/03/2020

\begin{itemize}
\item
On March 4, Governor Newsom declared a state of emergency after the
first death in California attributable to coronavirus occurred in
Placer County.

\item
\url{https://en.wikipedia.org/wiki/2020_coronavirus_pandemic_in_California}
\end{itemize}
\end{itemize}

\begin{itemize}
\item
School Closure (SC): 13/03/2020

\begin{itemize}
\item
On March 13, schools were closed in Marin, Sacramento, San Joaquin,
San Luis Obispo, Santa Clara, Solano, Placer, and Contra Costa
counties, as well as the Oakland, Antioch, Santa Cruz, Los Angeles
Unified, Chaffey Unified, Etiwanda, Fontana Unified,
Ontario-Montclair, Alta Loma Unified, San Diego, Los Alamitos Unified,
and Washington Unified school districts.

\url{https://en.wikipedia.org/wiki/2020_coronavirus_pandemic_in_California}
\end{itemize}
\end{itemize}

\begin{itemize}
\item
City Lockdown (CL): 19/03/2020

\begin{itemize}
\item
On March 19, Governor Newsom announced a statewide stay-at-home order.

\item
\url{https://en.wikipedia.org/wiki/2020_coronavirus_pandemic_in_California}
\end{itemize}
\end{itemize}

\item
Guangdong

\begin{itemize}
\item
School Closure (SC): 22/01/2020

\begin{itemize}
\item
School Winter Vocation in Guangdong was schedules as from 19/01/2020 to 16/02/2020. The semester start was delayed until the date of the data was collected (April 3).

\item
\url{http://edu.gd.gov.cn/zxzx/xwfb/content/post_2968669.html}
\end{itemize}
\end{itemize}

\begin{itemize}
\item
Government Advocation (CA): 23/01/2020:

\begin{itemize}
\item
On Jan 23, Guangdong launched Level 1 Response Mechanism.

\item
\url{http://www.gd.gov.cn/gdywdt/gdyw/content/post_2878901.html}
\end{itemize}
\end{itemize}

\begin{itemize}
\item
Contact Tracing (CT): 23/01/2020

\begin{itemize}
\item
On January 23, Guangdong Provincial Health Office required Guandong City Level
Health Office to find out close contacts of confirmed cases.

\item
\url{http://pmoedb2f8.pic35.websiteonline.cn/upload/af6v.pdf}
\end{itemize}
\end{itemize}

\begin{itemize}
\item
Mass Testing (MT): 23/01/2020

\begin{itemize}
\item
On January 23, Guangdong Provincial Health Office required Guandong City Level
Health Office to perform new Coronavirus nucleic acid testing for
close contacts.

\item
\url{http://pmoedb2f8.pic35.websiteonline.cn/upload/af6v.pdf}
\end{itemize}
\end{itemize}

\begin{itemize}
\item
City Lockdown (CL): 24/01/2020

\begin{itemize}
\item
In China, Chinese New Year Holiday was scheduled as from 24/01/2020 to 02/02/2020.
On January 28, Guangdong Province government announced that employees for essential business
will resume working on 10/02/2020. Others will gradually resume working upon situation evaluation.

\item
\url{http://www.gd.gov.cn/gdywdt/gdyw/content/post_2879851.html}

\url{http://www.gov.cn/zhengce/content/2020-01/27/content_5472352.html}
\end{itemize}
\end{itemize}

\begin{itemize}
\item
Mask Use (MU): 26/01/2020

\begin{itemize}
\item
On January 26, Guangdong Command Office for the Prevention and Control of the
outbreak of new coronavirus infected pneumonia announced that Guangdong residential
people are required to wear masks in public places.

\item
\url{http://wsjkw.gd.gov.cn/zwyw_gzdt/content/post_2879265.html}
\end{itemize}
\end{itemize}

\begin{itemize}
\item
International Travel Ban (ITB): 28/03/2020

\begin{itemize}
\item
On March 26, Chinese government decided to temporarily suspend entry of
foreigners with the current valid visa and residence permit for entry
into China from 0:00 on March 28, 2020.

\item
\url{https://www.fmprc.gov.cn/web/wjbxw_673019/t1761858.shtml}
\end{itemize}
\end{itemize}

\item
Hong Kong

\begin{itemize}
\item
Government Advocation (GA): 04/01/2020

\begin{itemize}
\item
On January 4, the Hong Kong government declared a "serious response level" to the virus
outbreak centred on Wuhan.

\item
\url{https://en.wikipedia.org/wiki/2020_coronavirus_pandemic_in_Hong_Kong}
\end{itemize}
\end{itemize}

\begin{itemize}
\item
Contact Tracing (CT): 04/01/2020

\begin{itemize}
\item
On January 4, the Hong Kong Government launched Preparedness and Response Plan
for Novel Infectious Disease of Public Health Significance, which
indicates that at serious response level, Department of Health will
put close contacts of confirmed cases of the novel infection under
quarantine/medical surveillance; and other contacts under medical
surveillance.

\item
\url{https://www.chp.gov.hk/files/pdf/govt_preparedness_and_response_plan_for_novel_infectious_disease_of_public_health_significance_eng.pdf}.
\end{itemize}
\end{itemize}

\begin{itemize}
\item
Mass Testing (MT): 04/01/2020

\begin{itemize}
\item
On January 4, Hong Kong Government launched Preparedness and Response Plan
for Novel Infectious Disease of Public Health Significance, which
indicates that at alert response level, Department of Health Conduct
laboratory testing for the novel pathogen as required, including
testing for Hospital Authority and private hospitals to enhance
laboratory surveillance on the novel pathogen.
\item
\url{https://www.chp.gov.hk/files/pdf/govt_preparedness_and_response_plan_for_novel_infectious_disease_of_public_health_significance_eng.pdf}
\end{itemize}
\end{itemize}

\begin{itemize}
\item
Mask use (MU): 08/01/2020

\begin{itemize}
\item
On January 8, Hong Kong's Centre for Health Protection (CHP) added
``Severe respiratory disease associated with a novel infectious agent"
to their list of notifiable diseases to expand their authority on
quarantine. The Hong Kong government also shortened hospital visits
and made it a requirement for visitors to wear face masks.

\item
\url{https://en.wikipedia.org/wiki/2020_coronavirus_pandemic_in_Hong_Kong}.
\end{itemize}
\end{itemize}

\begin{itemize}
\item
School Closure (SC): 22/01/2020

\begin{itemize}
\item
Most Hong Kong schoolchildren never returned to school following the
Chinese New Year holidays (from January 22 to February 1) after the government announced and then, and then the school closures were
extended to April 20.

\item
\url{https://www.theguardian.com/world/2020/mar/03/the-longest-holiday-parents-coping-with-coronavirus-school-closures-in-east-asia}

\url{https://hongkongfp.com/2020/03/17/coronavirus-hong-kong-school-closures-extended-cathay-adds-flights-students-returning-uk-us/}
\end{itemize}
\end{itemize}

\begin{itemize}
\item
International Travel Ban (ITB): 27/01/2020

\begin{itemize}
\item
On January 27, Mrs Carrie Lam announced that the Steering Committee cum Command Centre has decided to impose restrictions on all Hubei Province residents and people who visited the Hubei Province in the past 14 days from entering Hong Kong until further notice to reduce the chances of infected persons entering the city.
\item
\url{https://www.info.gov.hk/gia/general/202001/26/P2020012600713.htm}
\end{itemize}
\end{itemize}

\item
Hubei

\begin{itemize}
\item
Government Advocation (GA): 20/01/2020

\begin{itemize}
\item
On January 20, Wuhan City government formed specialized command for epidemic control (CEC)
to upgrade measures to cope with the epidemic including enhanced
protection over the medical workers.
\item
\url{https://en.wikipedia.org/wiki/2019-2020_coronavirus_pandemic_in_Hubei}).
\end{itemize}
\end{itemize}

\begin{itemize}
\item
School Closure (SC): 22/01/2020

\begin{itemize}
\item
School Winter Vocation in Hubei was scheduled from 15/01/2020 to 09/02/2020. The semester start was delayed until the date of the data was collected (April 3).
\item
\url{http://www.cnr.cn/hubei/yaowen/20191224/t20191224_524910001.shtml}
\end{itemize}
\end{itemize}

\begin{itemize}
\item
City Lockdown (CL): 23/01/2020

\begin{itemize}
\item
On January 23, 2020, the central government of the People's Republic
of China imposed a lockdown in Wuhan and other
cities.

\item
\url{https://en.wikipedia.org/wiki/2019-2020_coronavirus_pandemic_in_Hubei}
\end{itemize}
\end{itemize}

\begin{itemize}
\item
Mask Use (MU): 22/01/2020

\begin{itemize}
\item
On January 22, The city authority began to require all citizens to wear a mask in public places.
\item
\url{https://en.wikipedia.org/wiki/2019-2020_coronavirus_pandemic_in_Hubei}.
\end{itemize}
\end{itemize}

\begin{itemize}
\item
Public Transport Closure (PTC): 23/01/2020

\begin{itemize}
\item
On the early morning of January 23, the government of Wuhan City announced a lockdown at around 2 o'clock which said, ``Since 10:00 AM on January 23, 2020, the city's bus, metro, ferry, coach services will be suspended. Without a special reason, the residents should not leave Wuhan. Departure from the airport and railway stations will be temporarily prohibited. Recovery time of the services will be announced in a further notice."
\item
\url{https://en.wikipedia.org/wiki/2019-2020_coronavirus_pandemic_in_Hubei}
\end{itemize}
\end{itemize}

\begin{itemize}
\item
International Travel Ban (ITB): 23/01/2020

\begin{itemize}
\item
Tianhe International Airport, Wuhan's only civil airport suspended all
commercial flights from 13:00 on January 23.
\item
\url{https://en.wikipedia.org/wiki/2019-2020_coronavirus_pandemic_in_Hubei}
\end{itemize}
\end{itemize}

\begin{itemize}
\item
Contact Tracing (CT): 03/02/2020

\begin{itemize}
\item
In the New Coronavirus Prevention and Control Guideline (1st version), it was
indicated that the elderly who has been the close contacts with
suspected cases should be recorded, and schools should monitor the
health conditions of students and report to medical and health
institutions for possible tracing of close contacts.
\item
\url{http://wjw.hubei.gov.cn/bmdt/ztzl/yqxxfwxt/jkkp/202002/t20200202_2018199.shtml}
\end{itemize}
\end{itemize}

\begin{itemize}
\item
Mass Testing (MT): 05/02/2020

\begin{itemize}
\item
On February 5, a 2000-sq-meter emergency detection laboratory named
``Huo-Yan" was opened by BGI which can process over 10,000 samples a
day.
\item
\url{https://en.wikipedia.org/wiki/2019-2020_coronavirus_pandemic_in_Hubei}
\end{itemize}
\end{itemize}

\item
Macau

\begin{itemize}
\item
Government Advocation (GA): 31/12/2019

\begin{itemize}
\item
On December 31 2019, the Health Bureau was notified by the National
Health Commission of an outbreak of pneumonia of unknown cause in
Wuhan, Hubei. Residents of Macau were asked to avoid excessive panic but to be
conscious of personal hygiene and the hygiene of their
environment. Those traveling to Wuhan were advised to avoid visiting
local hospitals or having contact with sick people.
\item
\url{https://en.wikipedia.org/wiki/2020_coronavirus_pandemic_in_Macau}
\end{itemize}
\end{itemize}

\begin{itemize}
\item
School Closure (SC): 22/01/2020

\begin{itemize}
\item
On January 24, the Education and Youth Affairs Bureau announced that
all non-tertiary schools would extend their Chinese New Year holiday (from January 24 to January 29),
not resuming classes until February 10 or later. On January 30, the
Tertiary Education Bureau announced that the resumption of classes
would be delayed further, and that a schedule for resuming classes
would be released one week before classes were to resume.
\item
\url{https://en.wikipedia.org/wiki/2020_coronavirus_pandemic_in_Macau}.
\end{itemize}
\end{itemize}
\begin{itemize}
\item
International Travel Ban (ITB): 28/01/2020

\begin{itemize}
\item
On January 28, Secretary for Administration and Justice Zhang Yongchun said that in accordance with the decision of the central government, endorsements for mainland Chinese visitors to Macau would be suspended. Effective March 18, the government banned entry of all non-residents, with exceptions for mainland China, Hong Kong, and Taiwan.
\item
\url{https://en.wikipedia.org/wiki/2020_coronavirus_pandemic_in_Macau}
\end{itemize}
\end{itemize}

\begin{itemize}
\item
Mask Use (MU): 03/02/2020

\begin{itemize}
\item
On February 3, the government of Macau announced that starting at noon, all bus and taxi passengers were required to wear masks.
\item
\url{https://en.wikipedia.org/wiki/2020_coronavirus_pandemic_in_Macau}
\end{itemize}
\end{itemize}

\begin{itemize}
\item
Mass Testing (MT): 20/02/2020

\begin{itemize}
\item
On February 19, the government announced that effective February 20,
passengers coming from COVID-19 hotspots would need to undergo medical
checks upon entering Macau. Medical checks might also be conducted on
Macau residents who made multiple trips back and forth to Zhuhai every
day.
\item
\url{https://en.wikipedia.org/wiki/2020_coronavirus_pandemic_in_Macau}
\end{itemize}
\end{itemize}

\item
New York

\begin{itemize}
\item
International Travel Ban (ITB): 02/02/2020

\begin{itemize}
\item
On Jan 31, the United States imposed an entry ban on all foreign
nationals who were in the People's Republic of China, excluding
Taiwan, Hong Kong, and Macau, in the past fourteen days, effective
5:00 p.m. eastern standard time on February 2, 2020.
\item
\url{https://www.whitehouse.gov/presidential-actions/proclamation-suspension-entry-immigrants-nonimmigrants-persons-pose-risk-transmitting-2019-novel-coronavirus/}.
\end{itemize}
\end{itemize}

\begin{itemize}
\item
Government Advocation (GA): 07/03/2020

\begin{itemize}
\item
On March 7, Governor Andrew Cuomo declared a state of emergency.
\item
\url{https://en.wikipedia.org/wiki/2020_coronavirus_pandemic_in_New_York_(state)}
\end{itemize}
\end{itemize}

\begin{itemize}
\item
Mass Testing (MT): 13/03/2020

\begin{itemize}
\item
On March 13, drive-through testing began in New Rochelle, Westchester
County.
\item
\url{https://en.wikipedia.org/wiki/2020_coronavirus_pandemic_in_New_York_(state)}
\end{itemize}
\end{itemize}

\begin{itemize}
\item
School Closure (SC): 15/03/2020

\begin{itemize}
\item
On March 15, Governor Andrew Cuomo announced that New York City schools would close
the following day through April 20.
\item
\url{https://en.wikipedia.org/wiki/2020_coronavirus_pandemic_in_New_York_(state)}
\end{itemize}
\end{itemize}

\begin{itemize}
\item
School Closure (SC): 20/03/2020

\begin{itemize}
\item
On March 20, Governor Andrew Cuomo announced the statewide stay-at-home order
with a mandate that all non-essential workers work from
home.
\item
\url{https://en.wikipedia.org/wiki/2020_coronavirus_pandemic_in_New_York_(state)}
\end{itemize}
\end{itemize}

\item
Taiwan

\begin{itemize}
\item
Government Advocation (GA): 20/01/2020

\begin{itemize}
\item
On January 20, the government deemed the risk posed by the outbreak
sufficient to activate the Central Epidemic Command Center (CECC).
\item
\url{https://en.wikipedia.org/wiki/2020_coronavirus_pandemic_in_Taiwan}
\end{itemize}
\end{itemize}

\begin{itemize}
\item
School Closure (SC): 22/01/2020

\begin{itemize}
\item
Taiwan School Winter Vocation was scheduled from January 21 to February 11, which was further
extended to February 25.
\item
\url{https://en.wikipedia.org/wiki/2020_coronavirus_pandemic_in_Taiwan}
\end{itemize}
\end{itemize}

\begin{itemize}
\item
International Travel Ban (ITB): 23/01/2020

\begin{itemize}
\item
On January 23, Wuhan residents are banned to enter Taiwan. Starting from March 19,  foreign
nationals were barred from entering Taiwan, with some exceptions, such
as those carrying out the term of a business contract, holding valid
Alien Resident Certificates, diplomatic credentials, or other official
documentation and special permits.
\item
Wang, C. Jason, Chun Y. Ng, and Robert H. Brook. "Response to COVID-19 in Taiwan: big data analytics, new technology, and proactive testing." Jama (2020).
\item
\url{https://en.wikipedia.org/wiki/2020_coronavirus_pandemic_in_Taiwan}
\end{itemize}
\end{itemize}

\begin{itemize}
\item
Mask Use (MU): 31/01/2020

\begin{itemize}
\item
On March 31, transportation and communications minister Lin Chia-lung
announced that all passengers on trains and intercity buses were
required to wear masks, as were people at highway rest stops.
\item
\url{https://en.wikipedia.org/wiki/2020_coronavirus_pandemic_in_Taiwan}
\end{itemize}
\end{itemize}

\begin{itemize}
\item
Mass Testing (MT): 01/02/2020

\begin{itemize}
\item
On February 1, Taiwan CDC began monitoring all individuals who had
travelled to Wuhan within fourteen days and exhibited a fever or
symptoms of upper respiratory tract infections.
\item
\url{https://en.wikipedia.org/wiki/2020_coronavirus_pandemic_in_Taiwan}
\end{itemize}
\end{itemize}

\begin{itemize}
\item
Contact Tracing (CT): 27/01/2020

\begin{itemize}
\item
NHIA and NIA integrate patients's past 14-day travel history into NHIA database.
\item
Wang, C. Jason, Chun Y. Ng, and Robert H. Brook. "Response to COVID-19 in Taiwan: big data analytics, new technology, and proactive testing." Jama (2020).
\end{itemize}
\end{itemize}

\item
Washington

\begin{itemize}
\item
International Travel Ban (ITB): 02/02/2020

\begin{itemize}
\item
On January 31, the United States imposed an entry ban on all foreign
nationals who were in the People's Republic of China, excluding
Taiwan, Hong Kong, and Macau, in the past fourteen days, effective
5:00 p.m. eastern standard time on February 2, 2020.
\item
\url{https://www.whitehouse.gov/presidential-actions/proclamation-suspension-entry-immigrants-nonimmigrants-persons-pose-risk-transmitting-2019-novel-coronavirus/}
\end{itemize}
\end{itemize}

\begin{itemize}
\item
Government Advocation (GA): 29/02/2020

\begin{itemize}
\item
On February 29, Governor Jay Inslee declared a state of emergency.
\item
\url{https://en.wikipedia.org/wiki/2020_coronavirus_pandemic_in_Washington_(state)\#Government_response}
\end{itemize}
\end{itemize}

\begin{itemize}
\item
School Closure (SC): 13/03/2020

\begin{itemize}
\item
On March 12, Governor Inslee announced closures for all public and
private K-12 schools in King, Snohomish, and Pierce Counties beginning
from March 17 through at least April 24. On March 13, Inslee announced
K-12 closures until at least April 24 throughout the state.
\item
\url{https://en.wikipedia.org/wiki/2020_coronavirus_pandemic_in_Washington_(state)\#Government_response}
\end{itemize}
\end{itemize}

\begin{itemize}
\item
Mass Testing (MT): 16/03/2020

\begin{itemize}
\item
On March 16, the University of Washington has expanded its drive-through coronavirus testing to include first responders and University of Washington Medicine patients with COVID-19 symptoms, as well as the health care system's staff.
\item
\url{https://www.kuow.org/stories/drive-through-virus-testing-expands}
\end{itemize}
\end{itemize}

\begin{itemize}
\item
City Lockdown (CL): 23/03/2020

\begin{itemize}
\item
Governor Inslee announced a statewide stay-at-home order on March 23,
to last at least two weeks.
\item
\url{https://en.wikipedia.org/wiki/2020_coronavirus_pandemic_in_Washington_(state)\#Government_response}
\end{itemize}
\end{itemize}
\end{itemize}

\end{document}